\documentstyle[sprocl]{article}

\def\Journal#1#2#3#4{{#1} {\bf #2}, #3 (#4)}
 
\def\PLB{{\em Phys. Lett.}  B}
\def\PRL{\em Phys. Rev. Lett.}
\def\PRD{{\em Phys. Rev.} D}

\def\APPB{{\em Acta Phys. Polon.} B}

\def\AP{\em Ann. Physik}
 
\def\be{\begin{equation}}
\def\ee{\end{equation}}
\def\bea{\begin{eqnarray}}
\def\eea{\end{eqnarray}}
 
\begin{document}

\title{Flow equations in the light-front QCD: mass gap and confinement}

\author{E. GUBANKOVA}

\address{Department of Physics, North Carolina State University,\\ 
Raleigh, NC 27696-8202, USA\\E-mail: egoubank@unity.ncsu.edu}


\maketitle\abstracts{The light-front QCD is studied using 
the method of flow equations. Solving the light-front gluon gap equation,
the effective gluon mass is generated dynamically.
The effective interaction between static quark and antiquark,
generated through elimination of the quark-gluon minimal
coupling by flow equations, has the Coulomb,
$1/q^2$, and confining, $1/q^4$, singular behavior.
Elimination of the quark-gluon coupling at small gluon momenta
is governed by the cutoff dependent, dynamical gluon mass,
which makes this elimination possible 
and provides such an enhancement at $q\sim 0$.
The cutoff, which regulates small light-front $x$ divergences,
sets up a scale for the dynamical gluon mass and the string tension
in the $q\bar{q}$-potential.
The mechanism of confimenemet in the light-front frame is suggested,
based on the singular nature of the light-front gauge
along the light-front $x$-axis.}

\section{Introduction}

The perturbative aspects of non-abelian gauge theories
were underestood many years ago, and the perturbative calculations 
provided convincing proof that QCD is the theory of strong interactions. 
However nonperturbative QCD phenomena have been difficult  
to analyze mainly because calculational techniques are still lacking,
even though the qualitative features have been more or less
understood.

In particular, it is widely believed that pure Yang-Mills theory,
with no dynamical quarks, posseses the features
like asymptotic freedom, mass generation through the transmutation
of dimensions and confinement: linear potential between static (probe)
quarks. In the past few years there were several studies  
addressing the issue of confinement and generation of mass gap
in the framework of the Schr{\"o}dinger picture
\cite{DiakonovZarembo}, \cite{Karabali}. To mention a few early
works, see refs. \cite{Schr"odinger}. We have tried to understand these 
nonperturbative mechanisms also in a Hamiltonian framework,
solving flow equations for canonical QCD Hamiltonian  
in the light-front (LF) gauge selfconsistently for the few lowest sectors.
Dynamics of quarks has been excluded to disentangle
the complexity of chiral symmetry breaking.
Early studies of confinement in the LF frame
were conducted using similarity renormalization \cite{BrisudovaPerry}
and transverse lattice \cite{BvdSandeDalley},
based on the fact that QCD in $3+1$ already has a confining interaction
in the form of the instantaneous four-fermion interaction, 
which is the confining interaction in $1+1$. 
However, in our study the instantaneous interaction
is canceled by the corresponding term in the dynamical interaction,
generated by flow equations. The result is a three dimensional 
linear rising potential in the infrared region.

Our basic strategy has been to generate dynamically
an effective gluon mass (energy) through interaction 
with the LF 'vacuum' (exact zero modes on the light-front, $x=0$), 
and construct an effective interaction between 
static quarks as an exchange with this dynamical gluon, 
which forms a flux between quarks. One could construct 
a LF field theory with massive photons and gluons, 
as was done in \cite{Soper}, but this theory is in a Coulomb phase 
rather than confining because of the lack of a dynamical mass 
generation mechanism. Dynamical mass generation in 
the light-front frame dates back to the works of Cornwall \cite{Cornwall},       
where a kind of pinch technique was used to define
a gluon propagator and hence a gluon effective energy,
extracted from a Green's function for physical observables
(for example two-body scattering amplitude).      
It was suggested in \cite{Cornwall}, that in physical terms,
a gluon 'mass' may lead to a vortex condensation 
\footnote{In the massive gauge theory 
one needs to introduce a scalar field, which condenses, 
in order to preserve gauge invariance. Topologically  
this field represents vortices \cite{Cornwall}.
}
with $<Tr G_{\mu\nu}^2>\neq 0$, and conversely $<Tr G_{\mu\nu}^2>\neq 0$
generates a gluon mass. This provides a link with the existing
vortex picture of confinement. 

A subtle point in the LF field theory is that 
the LF vacuum is just empty space. 
Therefore it seems a problem how confinement can occur 
in the LF frame, and what quantity sets up a scale for a dynamical
mass gap and the string tension. We adopt a model, suggested 
by Susskind and Burkardt in the context of chiral symmetry breaking 
in the LF frame \cite{SusskindBurkardt}.  
In the parton model one pictures a fast moving hadron 
as being some collection of constituents with relatively large momentum,
such that when one boosts the system, doubles its momentum,
all these partons double their momenta and so forth.
Therefore one can formulate a field theory on the axis
of the light-front momentum $x$. 
Partons that form a hadron are at positive, finite $x$ and according
to Feynman and Bjorken fill the $x$-axis in a way which gets 
denser and denser as one goes to smaller $x$; and the vacuum is
at $x=0$. The fundamental property of LF Hamiltonians, 
that under a rescaling of the LF momentum,
$x\rightarrow \lambda x$, the LF Hamiltonian scales like
$H\rightarrow H/\lambda$, can be interpreted as a dilatation symmetry
along the $x$-axis, if one thinks of the $x$-axis as a spacial axis.
This symmetry holds on a classical level and it is brocken
on a quantum level by anomali.
As one approaches small $x$, interaction between partons gets stronger,
contributing divergent matrix elements. A natural cutoff
is provided by $\delta x=\varepsilon x$, 
a minimal spacing between constituents,
which plays the role of UV-regulator 
\footnote{Small $x$ correspond to the large light-front energies.}.
As long as $\delta x$ is finite, i.e. as long as the density of partons
on the $x$-axis is not infinite, one obtains finite matrix elements.
Cutoff $\delta x$ breaks the dilatation symmetry along the $x$-axis
and gauge symmetry, and sets up a scale for quantities 
of dimension of energy (mass) and higher powers in energy.
In particular, one generates dynamically an effective gluon mass, 
which in turn will define string tension between quark and antiquark.
Formation of the $q\bar{q}$ bound state through breaking an internal symmetry
can be viewed analogously to the creation of Cooper pairs in superconducter.

In terms of effective theory, the Hamiltonian below the LF cutoff,    
$H_{\leq \varepsilon}$, which desribes high energy (UV) strongly
correlated states, can be substituted by the v.e.v., since 
strongly coupled configurations are frozen. Hamiltonian above 
the cutoff $H_{ > \varepsilon}$ is treated by flow equations 
(renormalization group transformation).
One way of looking at the physics behind this v.e.v. 
and mass generation in gluodynamics is that the composite field $\phi$  
which creates $0^+$ glueballs has a finite v.e.v.
\cite{Cornwall}, \cite{Karabali}.

The dilatation symmetry, discussed above, reflects some underlying scale
invariance of the LF field theory formulated on $x$-axis. Introducing
the cutoff, breaks this symmetry. Because physics should remain
independent on the cutoff, one must be looking for a fixed point of 
the renormalization group. Therefore the right tool for studing 
such a system is the renormalization group, which is provided
by the method of flow equations \cite{Wegner}.

In the method of flow equations an effective interaction between quarks 
arises through elimination of quark-gluon coupling. 
Procedure converges in the UV for large gluon momenta transfer
$\vec{q}^{~2}$, but in the IR one is not able to eliminate vanishing 
gluon momenta, because the similarity factor
which governes the elimination does not decay for vanishing arguments.
In physical terms, one can not integrate soft gluons
in the same fashion. As soon as gluon aquires a dynamical mass,
which vanishes at large gluon momenta and not equal zero 
only at small momenta, the similarity factor decays even 
for small momenta transfer, because the argument contains
an effective gluon energy instead of only gluon mometum.
Therefore we can eliminate by flow equations even soft gluon modes,
that are responsible for the long-range part of an effective
quark potential. In the IR this gives $1/q^4$ behavior 
for an effective $q\bar{q}$ interaction.  

In the next section we solve flow equations for an effective gluon mass
and quark-antiquark effective interaction, based on the QCD Hamiltonian
in the light-front gauge.

\section{Gluon mass gap and an effective quark interaction}

Let $Q$ being a projection operator on a one-gluon state, and
$P$ on a $q\bar{q}$ state, $Q|\psi\rangle =|g\rangle$
and $P|\psi\rangle =|q\bar{q}\rangle$. Flow equations 
for matrix elements of the QCD Hamiltonian between these states read
\begin{eqnarray}
\frac{dE_q(l)}{dl} &=& -\sum_{p}\frac{1}{E_q(l)-E_p(l)}
\frac{d}{dl}\left( h_{qp}(l)h_{pq}(l) \right)
 \label{eq:1} \\
\frac{dh_{pp'}(l)}{dl} &=& -\sum_{q}\left(
\frac{dh_{pq}(l)}{dl}\frac{1}{E_p(l)-E_q(l)}h_{qp'}(l) +
h_{pq}(l)\frac{1}{E_{p'}(l)-E_q(l)}\frac{dh_{qp'}(l)}{dl}
\right) \nonumber
\,,\end{eqnarray}
where $p$ ($p'$) runs through all states in the $P$-subspace,
and $q$ in the $Q$-subspace. 
Flow equations for the quark-gluon coupling  $h_{pq}$
(coupling between $P$ and $Q$ sectors) are
\begin{eqnarray}
\frac{dh_{pq}(l)}{dl} &=& -\left(
E_p(l)-E_q(l) \right) \eta_{pq}(l)
\nonumber\\
\eta_{pq}(l) &=& -\frac{h_{pq}(l)}{E_p(l)-E_q(l)}\frac{d}{dl}
\left( \ln f(z_{pq}(l)) \right)
\nonumber\\
z_{pq}(l) &=& l\left( E_p(l)-E_q(l) \right)^2
\,,\label{eq:2}\end{eqnarray}
where $f$ is the similarity function with properties
$f(0) =1$ and $f(z\rightarrow\infty)=0$.
From Eq.~(\ref{eq:2}), elimination of quark-gluon coupling 
\begin{eqnarray}
h_{pq}(l)=h_{pq}(0)f(z_{pq}(l))
\,,\label{eq:3}\end{eqnarray} 
is governed by the similarity function $f(l(E_p(l)-E_q(l))^2)$, 
which vanishes for the matrix elements with energy differences 
exceeding the cutoff $\lambda$, $|E_p(l)-E_q(l)|\gg 1/\sqrt{l}=\lambda$,
$h_{pq}(0)$ is the initial value, and $E_q(l)$ ($E_p(l)$) is a solution of 
the flow (renormalization group) equation, Eq.~(\ref{eq:1}).
As long as the argument of the similarity function is not equal zero,
one can eliminate quark-gluon coupling and solve flow equations
for an effective gluon energy and effective $q\bar{q}$-interaction,
Eq.~(\ref{eq:1}). However, if one neglects the cutoff dependence
of energies, $E_p(l)\sim E_p(0)$,
the argument of the similarity function may become zero in the degenerate case,
$E_p(0)=E_q(0)$, and then the energy denominator blows up 
in effective interactions in Eq.~(\ref{eq:1}).
 
In physical terms, the leading behavior of the argument 
in the quark-gluon similarity function 
is given by an effective gluon energy transfer between quark and antiquark,
$\left(E_p(l)-E_q(l)\right)\sim E^{eff}_{gluon}(l)$, which reduces
at the initial value of flow parameter, $l=0$, to a gluon momentum transfer,
$E^{eff}_{gluon}(0)\sim \vec{q}^{~2}$.
Neglecting the cutoff dependence of the effective gluon energy,
one eliminates quark-gluon coupling only at large gluon momenta,
and fails at small gluon momenta, because the argument of the similarity function
tends to zero. For small gluon momenta the dependence of gluon effective energy
on the cutoff becomes important, given by a dynamically generated
gluon mass, that makes elimination of quark-gluon coupling possible
even for small gluon momenta. We therefore solve 
flow equations, Eq.~(\ref{eq:1}), consistently
for an effective gluon mass and an effective $q\bar{q}$-interaction below.
We show, that elimination of quark-gluon coupling
for high and low gluon momenta contribute correspondingly 
to the UV and IR parts of quark-antiquark potential,
that behave in momentum space like $1/\vec{q}^{~2}$ and $1/\vec{q}^{~4}$, 
respectively, where $\vec{q}$ is a gluon momentum transfer.

\subsection{Gluon gap equation}

In the light-front frame flow equation for an effective gluon mass
(the first equation in Eq.~(\ref{eq:1})),
with connection to the light-front energy 
$q^-=(q_{\perp}^2+\mu^2(\lambda))/q^+$, reads
\begin{eqnarray}
\frac{d\mu^2(\lambda)}{d\lambda} &=&
2T_fN_f\int_0^1\frac{dx}{x(1-x)}
\int_0^{\infty}\frac{d^2k_{\perp}}{16\pi^3}
g_q^2(\lambda)\frac{1}{Q^2_2(\lambda)}
\frac{df^2(Q^2_2(\lambda);\lambda)}{d\lambda}
\nonumber\\
&\times&
\left(
\frac{k_{\perp}^2+m^2}{x(1-x)}
-2k_{\perp}^2\right)
\nonumber\\
&+& 2C_a\int_0^1\frac{dx}{x(1-x)}
\int_0^{\infty}\frac{d^2k_{\perp}}{16\pi^3}
g_g^2(\lambda)\frac{1}{Q_1^2(\lambda)}
\frac{df^2(Q_1^2(\lambda);\lambda)}{d\lambda}
\nonumber\\
&\times&
\left(
k_{\perp}^2(1+\frac{1}{x^2}+\frac{1}{(1-x)^2})
\right)
\,,\label{eq:4}\end{eqnarray}
with
\begin{eqnarray}
Q_1^2(\lambda) &=& \frac{k_{\perp}^2+\mu^2(\lambda)}{x(1-x)}
-\mu^2(\lambda)
~,~
Q_2^2(\lambda) =
\frac{k_{\perp}^2+m^2}{x(1-x)}
-\mu^2(\lambda)
\,,\label{eq:5}\end{eqnarray}
where gluon couples to the quark-anti-quark pairs,
with the quark-gluon coupling $g_q(\lambda)$,
and pairs of gluons, with the three-gluon coupling $g_g(\lambda)$;
for finite $\lambda$ these couplings are different from each other 
functions of momenta; connection between flow parameter and the cutoff,
$l=1/\lambda^2$, is used; 
similarity function $f$ plays the role of UV regulator
in loop integrals; current quark mass is $m$; for $SU(N_c)$
$T_f\delta_{ab}={\rm Tr}(T^aT^b)=\frac{1}{2}\delta_{ab}$
and the adjoint Casimir is
$C_a\delta_{ab}=f^{acd}f^{bcd}=N_c\delta_{ab}$,
$N_c$ is the number of colors (i.e., $N_c=3$).

Coupled system of equations for the cutoff dependent, 
effective gluon and quark masses
was first derived in \cite{Glazek}. Here, neglecting the cutoff dependence
of quark mass this system is reduced to Eq.~(\ref{eq:4}).

Generally, it is difficult to solve Eq.~(\ref{eq:4}). One of the reasons
is that this equation contains (unknown) running coulings;
therefore the cutoff dependence of couplings is neglected below.
Second, initial condition for Eq.~(\ref{eq:4}) is not known.
The following renormalization condition to define the running gluon mass
$\mu(\lambda)$ through the 'physical' mass is assumed \cite{Glazek}:
the effective Hamiltonian at the scale $\lambda$ has bosonic eigenstates
with eigenvalues of the form $q^-=(q_{\perp}^2+\tilde{\mu}^2)/q^+$ 
\cite{Glazek}, i.e.  
\begin{eqnarray}
\frac{q_{\perp}^2+\tilde{\mu}^2}{q^+}\langle q',q\rangle &=&
\frac{q_{\perp}^2+\mu^2(\lambda)}{q^+}\langle q',q\rangle
\nonumber\\
&-& \int^{\lambda} d\lambda' \sum_{p}
\left(\eta_{q'p}(\lambda')h_{pq}(\lambda')
-h_{q'p}(\lambda')\eta_{pq}(\lambda')\right) 
\,,\label{eq:6}\end{eqnarray}
where $\tilde{\mu}$ denotes the 'physical' gluon mass;
the generator $\eta_{pq}$, given by Eq.~(\ref{eq:2}), eliminates
the quark-gluon (three-gluon) coupling $h_{pq}$;
$|q\rangle$ denotes a single effective gluon state with momentum
$q^+$ and $q_{\perp}$, 
$\langle q'|q\rangle =16\pi^3q^+\delta^{(3)}(q'-q)$.
Neglecting the cutoff dependence of the gluon mass on the r.h.s. 
of Eq.~(\ref{eq:4}),
the renormalization point is choisen at $q^2=\tilde{\mu}^2$
(for in- and out-going and intermediate state gluons).
Third, severe divergences arise at small light-front momenta $x$
and must be regularized. For this purpose,
as suggested by Zhang and Harindranath in \cite{ZhangHarindranath},
the minimum cutoff $u$ for transverse momentum $k_{\perp}$ is introduced.
Therefore, flow equation, Eq.~(\ref{eq:4}), is integrated in the finite limits,
$[u;\lambda]$. Explicitly, from the similarity function in gluon loop 
(the second term in Eq.~(\ref{eq:4})) one has
$u\leq\tilde{Q}_1^2=(k_{\perp}^2+\tilde{\mu}^2)/x(1-x)-\tilde{\mu}^2\leq\lambda$,
that restricts the transverse momentum to
$k_{\perp min}=(u^2+\tilde{\mu}^2)x(1-x)-\tilde{\mu}^2\leq k_{\perp}\leq
k_{\perp max}=(\lambda^2+\tilde{\mu}^2)x(1-x)-\tilde{\mu}^2$,
and the light-front momentum to
$\tilde{\mu}^2/(u^2+\tilde{\mu}^2)\leq x\leq 1-\tilde{\mu}^2/(u^2+\tilde{\mu}^2)$.
In the same way one finds restrictions on mometa in the quark-gluon loop 
(the first term in Eq.~(\ref{eq:4})).
Integrating Eq.~(\ref{eq:4}) with all constraints discussed above
and assuming the condition of Eq.~(\ref{eq:6}), gives \cite{Gubankova}
\begin{eqnarray}
\mu^2(\lambda) &=& 
\tilde{\mu}^2+\delta\mu_{PT}^2(\lambda)
+\delta\mu_{dyn}^2(\lambda,\tilde{\mu},u)
\,.\label{eq:7}\end{eqnarray}
where the perturbative term
\begin{eqnarray}
\delta\mu_{PT}^2(\lambda)=\frac{g^2}{4\pi^2}
\lambda^2 \left(
C_a (\ln\frac{u^2}{\tilde{\mu}^2}-\frac{11}{12})
+T_fN_f\frac{1}{3}\right)
\,.\label{eq:8}\end{eqnarray}
reproduces the result of the LF perturbation theory,
with renormalization point $q^2=0$, \cite{ZhangHarindranath}.
When renormalization is performed in the second oder,
the perturbative mass correction is absorbed by the mass counterterm,
$m_{CT}^2=-\delta\mu_{PT}^2(\lambda=\Lambda\rightarrow\infty)$.
After the perturbative renormalization is completed,
the dynamical mass
\begin{eqnarray}
\mu_{dyn}^2(\lambda) &=& \tilde{\mu}^2+\delta\mu_{dyn}^2(\lambda,\tilde{\mu},u)
=\tilde{\mu}^2+\sigma(\tilde{\mu},u)\ln\frac{\lambda^2}{\tilde{\mu}^2}
\nonumber\\
\sigma(\tilde{\mu},u) &=& -\frac{g^2}{4\pi^2}
\tilde{\mu}^2 \left( C_a (-\frac{u^2}{\tilde{\mu}^2}
+\ln\frac{u^2}{\tilde{\mu}^2}-\frac{5}{12})
+T_fN_f(\frac{1}{3}+\frac{m^2}{\tilde{\mu}^2}) \right)
\,,\label{eq:9}\end{eqnarray}
is left. Note, that the gluon mass $\mu_{dyn}$ 
is generated dynamically by flow equations.  
In the limit $\tilde{\mu}\rightarrow 0$,
one has $\sigma=\lim_{\tilde{\mu}\to 0}\sigma(\tilde{\mu},u)
=u^2g^2C_a/2\pi^2$, and, as shown below, 
$\sigma$ plays the role of the string tension between quark and antiquark. 
Scale $u$ is introduced by regulating the divergences 
at small LF momenta, $x\sim 0$, (small LF $x$ regularization). 
The value $u$ sets up a scale for the dynamical gluon mass and the string tension.

\subsection{Effective $q\bar{q}$-interaction}

An effective interaction, generated between quark and antiquark
by flow equations in the LF frame
(the second equation in Eq.~(\ref{eq:1})), reads   
\begin{eqnarray}
V_{q\bar{q}}=-Const\langle\gamma^{\mu}\gamma^{\nu}\rangle
\lim_{\tilde{\mu}\to 0} B_{\mu\nu}
\,,\label{eq:10}\end{eqnarray}
where instead of the strong coupling $\alpha_s$ some constant, $Const$,
is introduced; $\langle\gamma^{\mu}\gamma^{\nu}\rangle$
is a current-current term in the exchange channel.
At the end we set the gluon mass renormalization point
('physical' gluon mass) to zero, $\tilde{\mu}\to 0$;
thus eliminating the sensitivity to the renormalization point.
The kernel includes dynamical interaction and instantaneous exchange
with nonzero dynamical gluon mass,
given by
\begin{eqnarray}
B_{\mu\nu}=g_{\mu\nu}
\left(I_1+I_2\right)
+\eta_{\mu}\eta_{\nu}
\frac{\delta Q^2}{q^{+2}}
\left(I_1-I_2\right)
\,,\label{eq:11}\end{eqnarray}
where $g_{\mu\nu}$ is the LF metric tensor, and the LF vector $\eta_{\mu}$
is defined as $\eta\cdot k=k^+$.
The cutoff dependence of four-momentum transfers along 
quark and antiquark lines is accumulated in the factor 
\begin{eqnarray}
I_1=\int_0^{\infty}d\lambda\frac{1}{Q_1^2(\lambda)}
\frac{df(Q_1^2(\lambda);\lambda)}{d\lambda}
f(Q_2^2(\lambda);\lambda)
\,,\label{eq:12}\end{eqnarray}
with mometum transfers defined in the light-front frame as
\begin{eqnarray}
Q_1^2(\lambda) &=& Q_1^2+\mu^2_{dyn}(\lambda)
~,~
Q_1^2 = \frac{(x'k_{\perp}-xk'_{\perp})^2+m^2(x-x')^2}{xx'}
\nonumber\\
Q_2^2(\lambda) &=& Q_2^2+\mu^2_{dyn}(\lambda)
~,~
Q_2^2 = Q_1^2|_{x\rightarrow (1-x);~x'\rightarrow (1-x')}
\,,\label{eq:13}\end{eqnarray} 
and the dynamical gluon mass $\mu_{dyn}$ is given by Eq.~(\ref{eq:9}); 
also the following momenta are used
$Q^2=(Q_1^2+Q_2^2)/2$ and $\delta Q^2=(Q_1^2-Q_2^2)/2$.
Calculating Eq.~(\ref{eq:10}) with an explicit form of similarity function,
gives in the leading order $\delta Q^2\ll Q^2$
the following $q\bar{q}$-interaction \cite{Gubankova}
\begin{eqnarray}
V_{q\bar{q}}=-\langle\gamma^{\mu}\gamma_{\mu}\rangle
\left(C_f\alpha_s\frac{4\pi}{Q^2}+\sigma\frac{8\pi}{Q^4}
\right) 
\,,\label{eq:14}\end{eqnarray}
where to this order $Q^2$ reduces in the instant frame to
the square of gluon momentum transfer, $Q^2\sim \vec{q}^{~2}$,
with the gluon momentum $\vec{q}=(q_z,q_{\perp})$. 
In order to reproduce the standard Coulomb and linear confining potentials,
the correct prefactors are restored, using the freedom to fit 
$Const$ and $\sigma$ terms. 
Confining term in Eq.~(\ref{eq:14}), with singular behavior like 
$1/\vec{q}^{~4}$, arise from the elimination of the quark-gluon coupling
at small gluon momenta, that is governed by the cutoff dependent, 
dynamical gluon mass.   

\section*{Acknowledgments}
The author would like to thank the organizers of the workshop
for hospitality and support. The author is thanful to Stan Brodsky
and Lev Lipatov for helpful discussions. 
This work was supported 
by DOE grants DE-FG02-96ER40944, DE-FG02-97ER41048 and DE-FG02-96ER40947.

\end{document}